\documentclass[a4paper,preprintnumbers,floatfix,superscriptaddress,prl,twocolumn,showpacs,longbibliography]{revtex4-1}

\usepackage[utf8]{inputenc}
\usepackage[T1]{fontenc}
\usepackage[sc,osf]{mathpazo}
\usepackage{amsmath, amsthm, amssymb,amsfonts}
\usepackage{graphicx}
\usepackage{dcolumn}
\usepackage{bm}
\usepackage{bbm}
\usepackage{hyperref}
\usepackage{color}
\usepackage{siunitx}

\def\1{\mathbf{1}}
\def\0{\mathbf{0}}

\newcommand{\ket}[1]{| #1 \rangle}

\newcommand{\beq}{\begin{equation}}
\newcommand{\eeq}{\end{equation}}
\newcommand{\bea}[1]{\begin{equation}\begin{array}{#1}}
\newcommand{\eea}{\end{array}\end{equation}}
\newcommand{\beqn}{\begin{eqnarray}}
\newcommand{\eeqn}{\end{eqnarray}}

\renewcommand{\rho}{\varrho}

\newcommand{\processnext}[1]{%
  \ifx\listfinish#1\empty\else\listact{#1}\expandafter\processnext\fi}

\newcommand{\figref}[1]{Fig.~\ref{#1}}

\newcommand{\ea}{\end{eqnarray}}
\newcommand{\ban}{\begin{eqnarray*}}
\newcommand{\ean}{\end{eqnarray*}}

{\begin{framed}\begin{small}}
{\end{small}\end{framed}}

\DeclareGraphicsExtensions{.pdf,.png,.jpg}

\makeindex

\begin{document}
\title{Experimental non-locality in a quantum network}
\date{\today}

\author{Gonzalo Carvacho}
\affiliation{Dipartimento di Fisica - Sapienza Universit\`{a} di Roma, P.le Aldo Moro 5, I-00185 Roma, Italy}

\author{Francesco Andreoli}
\affiliation{Dipartimento di Fisica - Sapienza Universit\`{a} di Roma, P.le Aldo Moro 5, I-00185 Roma, Italy}

\author{Luca Santodonato}
\affiliation{Dipartimento di Fisica - Sapienza Universit\`{a} di Roma, P.le Aldo Moro 5, I-00185 Roma, Italy}

\author{Marco Bentivegna}
\affiliation{Dipartimento di Fisica - Sapienza Universit\`{a} di Roma, P.le Aldo Moro 5, I-00185 Roma, Italy}

\author{Rafael Chaves}
\affiliation{International Institute of Physics, Federal University of Rio Grande do Norte, 59070-405 Natal, Brazil}
\affiliation{Institute for Theoretical Physics, University of Cologne, 50937 Cologne, Germany}

\author{Fabio Sciarrino}
\email{fabio.sciarrino@uniroma1.it}
\affiliation{Dipartimento di Fisica - Sapienza Universit\`{a} di Roma, P.le Aldo Moro 5, I-00185 Roma, Italy}

\begin{abstract}
Non-locality stands nowadays not only 
as one of the cornerstones of quantum theory, but also plays a crucial role in quantum information processing. Several experimental investigations of non-locality have been carried out over the years. In spite of their fundamental relevance, however, all previous experiments do not consider a crucial ingredient that is ubiquitous in quantum networks: the fact that correlations between distant parties are mediated by several, typically independent, sources of quantum states. Here, using a photonic setup we investigate a quantum network consisting of three spatially separated nodes whose correlations are mediated by two independent sources. This scenario allows for the emergence of a new kind of non-local correlations that we experimentally witness by violating a novel Bell inequality. Our results provide the first experimental proof-of-principle of generalizations of Bell's theorem for networks, a topic that has attracted growing attention and promises a novel route for quantum communication protocols.
\end{abstract}

\maketitle

As demonstrated by the celebrated Bell's theorem \cite{Bell1964}, correlations arising from experiments with distant quantum mechanical systems are at odds with one of our most intuitive scientific notions, that of local realism. The assumption of realism formalizes the idea that physical quantities have well defined values independently of whether they are measured or not. In turn, local causality posits that correlations between distant particles can only originate from causal influences in their common past. Strikingly, these two rather natural assumptions together imply strict constraints on the empirical correlations that are compatible with them. These are the famous Bell inequalities, which have been recently violated in a series of loophole-free experiments \cite{Hensen2015,Shalm2015,Giustina2015} and thus
conclusively established the phenomenon known as quantum non-locality \cite{Brunner2014}. Apart from their profound implications in our understanding of nature, such experiments provide a proof-of-principle for practical applications of non-locality, most notably in the context of quantum networks \cite{Acin2007,Kimble2008,Sangouard2011}.

\begin{figure}[!t]
\includegraphics[width=0.98\columnwidth]{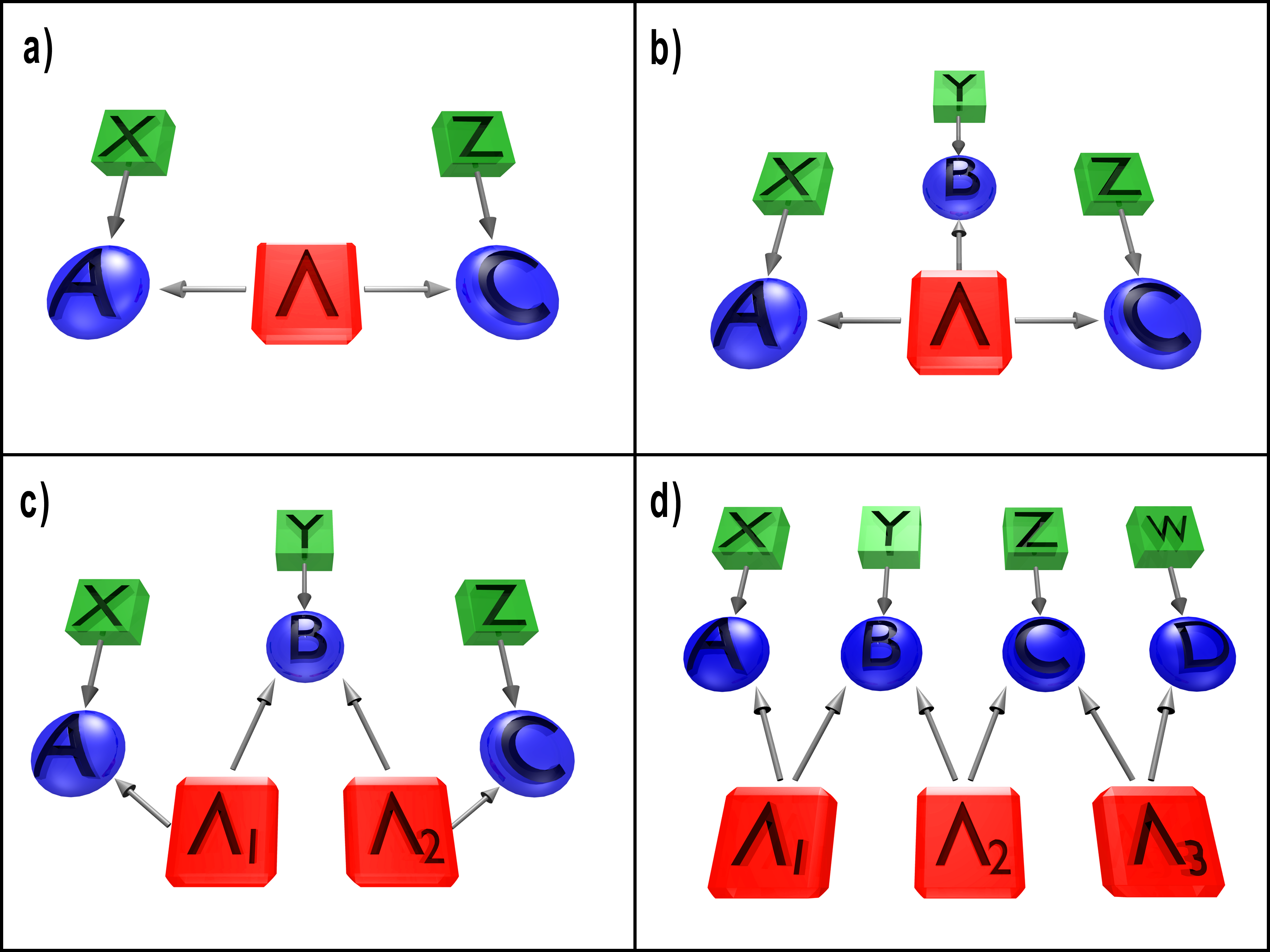}
\caption{{\bf Representation of the causal structures underlying the networks as directed acyclic graphs} \cite{Pearlbook}. Nodes in the graph represent the relevant random variables in the network and the arrows account for their causal relations. {\bf a)} Bipartite LHV model.  {\bf b)} Tripartite LHV model. {\bf c)} Tripartite scenario with two independent local hidden variables, i.e. BLHV model. {\bf d)} Possible extension of the bilocal model to a linear chain of four stations with three independent local hidden variables.
}
\label{fig:lhv}
\end{figure}

In a quantum network, short-distance nodes are connected by sources of entangled systems which can, via an entanglement swapping protocol \cite{Zukowski1993}, establish entanglement across long distances as well. Importantly, such long-distance entanglement can in principle also be used to violate a Bell inequality and thus establish a secure communication channel \cite{Ekert1991,Barrett2005,Vazirani2014}. Clearly, for these and many other potential applications \cite{Buhrman2010,Pironio2010,Colbeck2012,Reichardt2013}, the certification of non-local correlations across the network will be crucial. The problem, however, resides on the fact that experimental imperfections accumulate very rapidly as the size of the network and the number of sources of states increase, making the detection of non-locality very difficult or even impossible by usual means \cite{Sen2005,Cavalcanti2011}. One of the difficulties stems from the derivation of Bell inequalities themselves, where it is implicitly assumed that all the correlations originate at a single common source (see \figref{fig:lhv}-b), the so-called \textit{local hidden variable} (LHV) models. Notwithstanding, in a network a precise description must take into account that there are several and independent sources of states (see \figref{fig:lhv}-c), which introduce additional structure to the set of classically allowed correlations. In fact, there are quantum correlations that can emerge in networks that, while admitting a LHV description, are incompatible with any classical description where the independence of the sources is considered \cite{Fritz2012,Branciard2010,Branciard2012,Tavakoli2014,Chaves2015a,Chaves2016,Rosset2016}. That is, such networks allow for the emergence of a new kind of non-local correlations.

\begin{figure*}[ht!]
    \includegraphics[width=\textwidth]{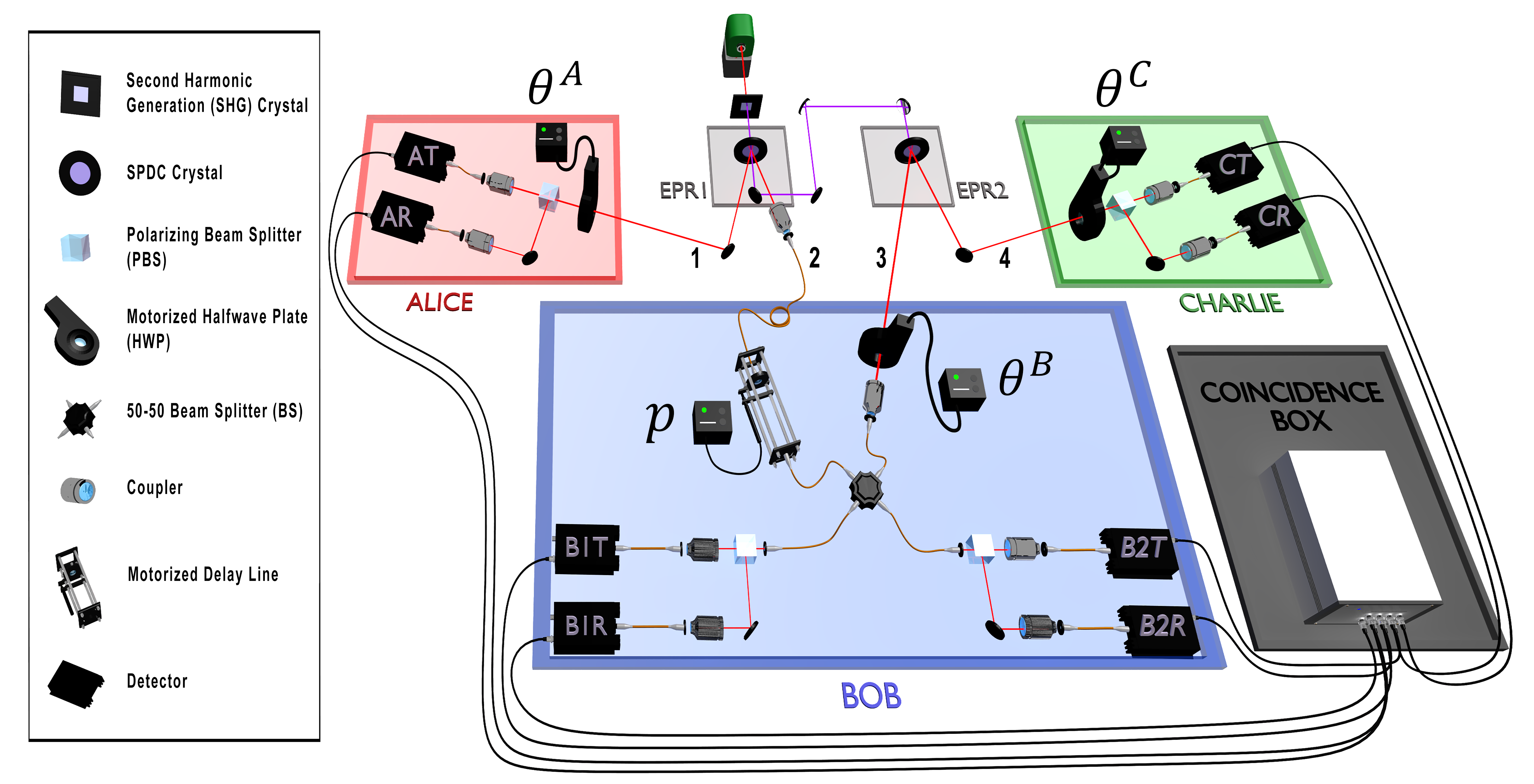}
\caption{{\bf Experimental apparatus for the violation of bilocality.} Two polarization-entangled photon pairs are generated via Spontaneous Parametric Down-Conversion (SPDC) in two separated non-linear crystals. Photon 1 (4) of the first (second) pair is directed to Alice's (Charlie's) station, where one of the local observables $A_0, \;A_1$ ($C_0, \;C_1$) is measured via a motorized Half Wave Plate (HWP) (angles $\theta^A$ and $\theta^C$) followed by a Polarizing Beam Splitter (PBS). Photons 2 and 3 are sent to Bob's station, where a complete Bell-state measurement is performed. A $50/50$ in-fiber BS followed by two PBSs allows to discriminate $\ket{\Psi^-}$ and $\ket{\Psi^+}$ when the HWP angle $\theta^B$ is set to zero and discriminate $\ket{\Phi^-}$ and $\ket{\Phi^+}$ when $\theta^B=45\si{\degree}$. A motorized delay line is adopted to control the amount of noise $p$ in the Bell measurement, by changing the photons wavepacket temporal overlap in the BS.}
\label{setup}
\end{figure*}

The aim of this paper is to experimentally observe, for the first time, this new type of non-locality. We experimentally implemented, using pairs of polarization-entangled photons, the simplest possible quantum network akin to a three-partite entanglement swapping scheme (see \figref{fig:lhv}-c). Two distant parties, Alice and Charlie, perform analysis measurements over two photons (1 and 4, see Fig. \ref{setup}) which were independently generated in two different sources, while a third station, Bob, performs a Bell-state measurement over the two other photons (2 and 3), one entangled with Alice's photon and the other entangled with Charlie's one. For sufficient low noise, upon conditioning on Bob's measurement outcome one can generate entanglement and non-local correlations between the two remaining particles, even though they have never interacted \cite{Zukowski1993}. Passed a certain noise threshold, however, no non-local correlations can be extracted from the swapped quantum state even though the correlations on the entire network might still display non-locality. We prove that this is indeed the case by violating a novel Bell inequality proposed in \cite{Branciard2010}. Further, showing that our experimental data is nevertheless compatible with usual LHV models where the independence of the sources is not taken into account, we can conclude that the non-local correlations we generate across the network are truly of a new kind.

Before entering the details and results of our experiment we start describing the typical scenario of interest in the study of quantum non-locality shown in \figref{fig:lhv}-b for the case of three distant parties. A source distributes a physical system to each of the parties that at each run of the experiment can perform the measurement of different observables (labelled by $x$, $y$ and $z$) thus obtaining the corresponding measurement outcomes (labelled by $a$, $b$ and $c$). In a classical description of such experiment, no restrictions other than local realism are imposed, meaning that the measurement devices are treated as black-boxes that take random (and independently generated) classical bits as inputs and produce classical bits as outputs as well. After a sufficient number of experimental runs is performed, the probability distribution of their measurements can be estimated, that according to the assumption of local realism can be decomposed as a LHV model of the form
\begin{eqnarray}
\label{LHV_tripartite}
p(a,b,c \vert x,y,z)= \sum_{\lambda} p(\lambda) p(a \vert x,\lambda)p(b \vert y,\lambda)p(c \vert z,\lambda).
\end{eqnarray}
The hidden variable $\lambda$ subsumes all the relevant information in the physical process and thus includes the full description of the source producing the particles as well as any other relevant information for the measurement outcomes.

In the description of the LHV model \eqref{LHV_tripartite} no mention is made about how the physical systems have been produced at the source. For the network we consider here (see \figref{fig:lhv}-c), the two sources produce states independently, thus the set of classically allowed correlations
\begin{eqnarray}
\label{LHV_bilocal}
p(a,b,c \vert x,y,z) = & & \sum_{\lambda_1,\lambda_2} p(\lambda_1)p(\lambda_2)  \\ \nonumber
& & p(a \vert x,\lambda_1)p(b \vert y,\lambda_1,\lambda_2)p(c \vert z,\lambda_2),
\end{eqnarray}
is now mediated via two independent hidden variables $\lambda_1$ and $\lambda_2$ \cite{Branciard2010}, thus defining a \textit{bilocal hidden variable} (BLHV) model.

In our setup, Bob always performs the same measurement (no measurement choice) obtaining four possible outcomes that can be parameterized by two bits $b_0$ and $b_1$. Alice and Charlie can choose each time one of two possible dichotomic measurements. Thus, in this case the observable distribution containing the full information of the experiment is given by $p(a,b_0,b_1,c \vert x,z)$. This allows us to violate the bilocality inequality proposed in \cite{Branciard2010} and further developed in \cite{Branciard2012,Tavakoli2014,Chaves2015a,Chaves2016,Rosset2016}:
\begin{equation}
\label{bilocalineq}
\mathcal{B}=\sqrt{|I|}+\sqrt{|J|} \leq 1
\end{equation}
The terms $I$ and $J$ are sums of expectation values, given by $\textstyle I=\frac{1}{4} \textstyle \sum\nolimits_{x,z} \langle A_{x}B_{0}C_{z}\rangle$ and $J=\frac{1}{4} \textstyle \sum\nolimits_{x,z} (-1)^{x+z} \langle A_{x}B_{1}C_{z}\rangle$
where $\textstyle \langle A_{x}B_{y}C_{z} \rangle= \textstyle \sum\nolimits_{a,b_{0},b_{1},c} (-1)^{a+b_{y}+c} p(a,b_{0},b_{1},c|x,z)$ and $x,z,a,b_0,b_1,c=0,1$. Inequality \eqref{bilocalineq} is valid for any classical model of the form \eqref{LHV_bilocal} and its violation demonstrates the non-local character of the correlations we produce among the network. \\

We generate entangled photon pairs via type-II Spontaneous Parametric Down-Conversion process (SPDC) occurring in two separated nonlinear crystals ($EPR\;1$ and $EPR\;2$) injected by a pulsed pump laser  (see Fig. \ref{setup}). When a pair of photons is generated in each of the crystals, one photon from source $EPR\;1$ ($EPR\;2$) is sent to Alice's (Charlie's) measurement station, where  polarization analysis in a basis which can be rotated of an arbitrary angle $\theta^A$ ($\theta^C$) is performed. The other two photons (2 and 3) are sent to Bob's station, which consists of an in-fiber 50/50 Beam Splitter (BS) followed by two Polarizing Beam Splitters (PBS) for the polarization analysis of each of the outputs. In the ideal case (which relies on perfect photons' indistinguishability), an incoming $\ket{\Psi^-}$ (singlet) state will feature antibunching, giving rise to coincidence counts at different outputs of the BS. All the other cases (triplet states) will experience bosonic bunching, ending up in the same BS output. A twofold coincidence corresponding to different polarizations in a single BS output branch corresponds to $\ket{\Psi^+}$ detection. A Half Wave Plate (HWP) placed before one of the arms of the BS allows, by setting $\theta^B=45\si{\degree}$, to change the incoming state from $\ket{\Phi^+}$ to $\ket{\Psi^-}$ and from $\ket{\Phi^-}$ to $\ket{\Psi^+}$. In this way, depending on the setting $\theta^B$, we are able to detect either $\ket{\Psi^+}$ and $\ket{\Psi^-}$, or $\ket{\Phi^+}$ and $\ket{\Phi^-}$ states.

By this approach, measuring all the combinations $(A_0, C_0), (A_0, C_1), (A_1, C_0), (A_1, C_1)$ of the observables $A_{0}=C_{0}=(\sigma_{z}+\sigma_{x})/\sqrt{2}$ and $A_{1}=C_{1}=(\sigma_{z}-\sigma_{x})/\sqrt{2}$ of Alice and Charlie, for the two possible $\theta^B$ configurations, we are able to reconstruct the probability $p(a,b_0,b_1,c \vert x,z)$ and then to compute the quantities $I$ and $J$ which appear in \eqref{bilocalineq}. The maximum value reached in our experimental setup was $\mathcal{B}=1.268\pm0.014$, corresponding to a violation of inequality \eqref{bilocalineq} of almost 20 sigmas. This value is fully compatible with a theoretical model that considers both colored and white noise in the state generated by the SPDC sources and takes into account the partial distinguishability of the generated photons (see Supplementary Information).

\begin{figure*}[ht!]
 \includegraphics[width=\textwidth]{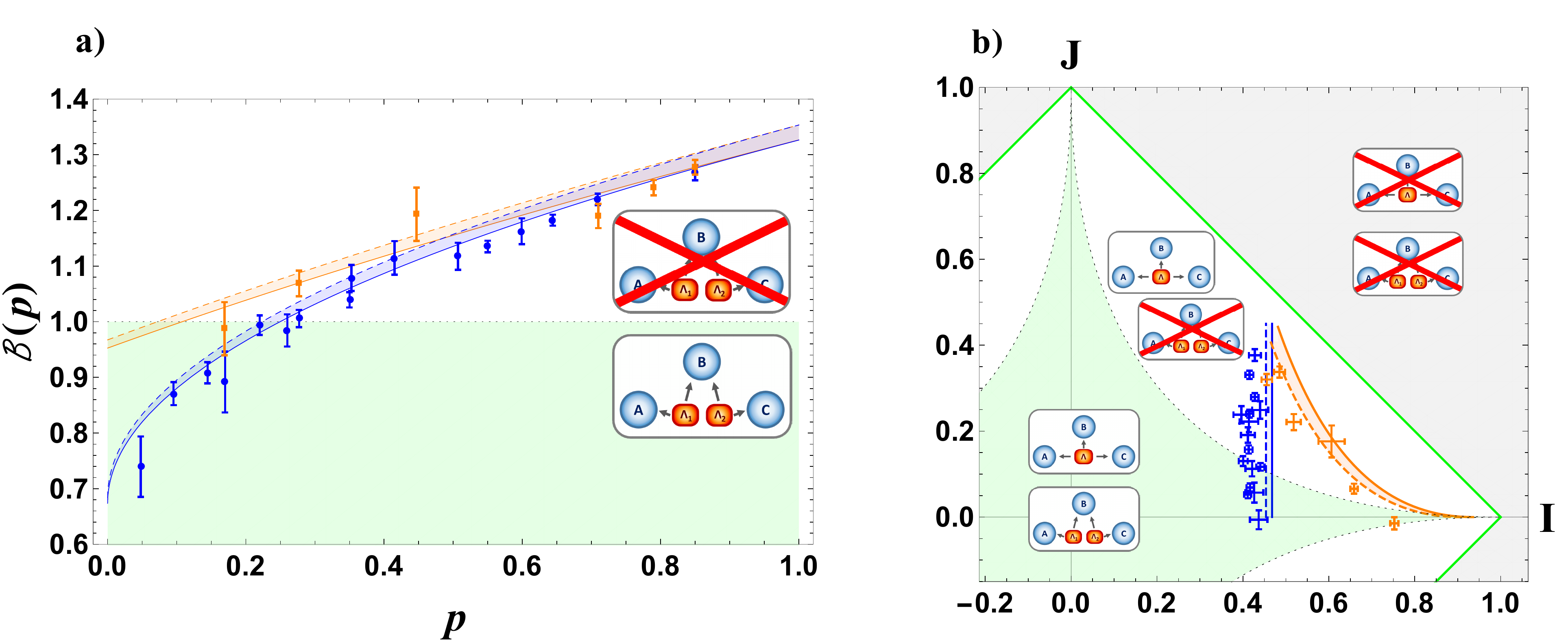}
 \caption{{\bf Experimental violation of bilocality.} {\bf a)} Measured quantity $\mathcal{B}$ as a function of the noise parameter $p$, with fixed (blue circles) and optimized (orange squares) measurements settings.
Theoretical predictions are shown by blue and orange shaded regions compatible with our state preparation and varying the other noise parameter $p$. The regions are obtained considering the propagation of the uncertainty in the experimental estimation of noises and are bounded by one standard deviation upper (dashed) and lower (line) curves. The dotted horizontal line indicates the bound of the inequality \eqref{bilocalineq}, while error bars indicates one standard deviation of uncertainty, due to Poissonian statistics.
{\bf b)}  Measured values in the $I$-$J$ plane. Error bars show one standard deviation for both $I$ and $J$ values. The dashed line bounds the bilocal region as prescribed by inequality \eqref{bilocalineq}. Green lines define the local set and the white area represents correlations which are compatible with local models but incompatible with bilocality assumption. The grey area shows the set of correlations which are incompatible with both local and bilocal models.
} \label{Results}
\end{figure*}

Next we address the robustness of the bilocality inequality violation with respect to experimental noise. To this aim, we tuned the noise in the Bell-state measurement by modifying the temporal overlap between photons 2 and 3. This can be achieved by using a delay line before one of the two inputs of the BS, thus controlling the temporal delay between these photons (see \figref{setup}). We can therefore define a noise parameter $p$ which is equal to $1$ in the ideal case of a perfect Bell-state measurement and is equal to $0$ when the probability of having a successful measurement is $1/2$. This parameter can be tuned from $p_{max}$ to zero by changing the delay from zero to a value larger than the coherence time of the photons.

The measured values of $\mathcal{B}$ versus $p$ are shown in Fig. \ref{Results}-a. As expected the violation decreases with increasing noise \cite{Branciard2010,Branciard2012}. This plot shows two sets of different data points: considering a fixed measurement basis (optimal in the absence of the additional noise) and optimizing the measurement basis at Alice and Charlie's stations as a function of $p$, i.e. changing the measurement basis in order to counteract noise effects. In both cases our setup can tolerate a substantial amount of noise before inequality \eqref{bilocalineq} is not violated anymore, but it is clear how the optimization increases both the degree and region of bilocality violation.

\begin{figure*}[ht!]
    \includegraphics[width=\textwidth]{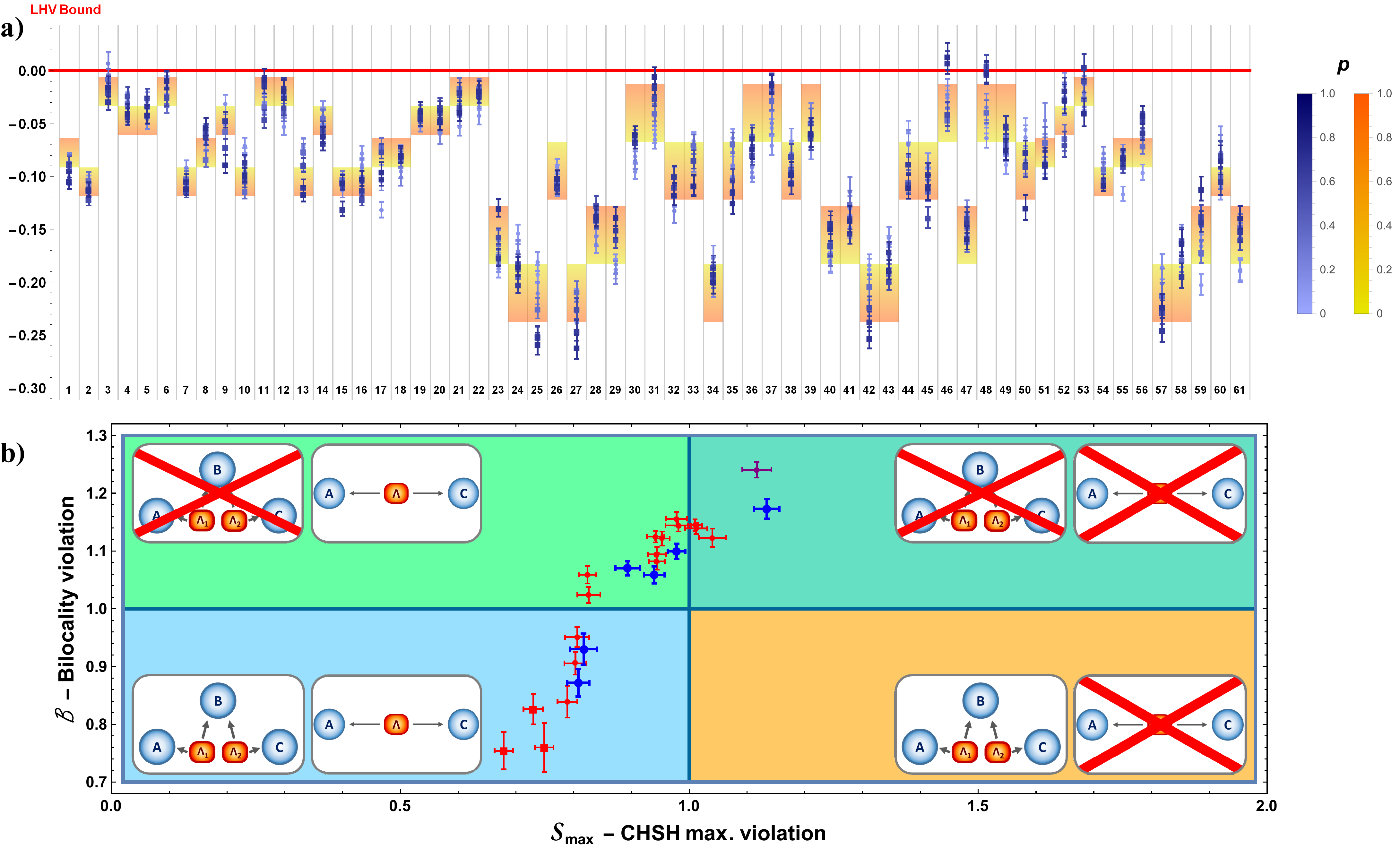}
 \caption{{\bf Experimental test of LHV models.} {\bf a)} Experimental violation values for all the 61 Bell inequalities compatible with a LHV model. Each column corresponds to a different inequality, rescaled in a way that a resulting value greater than $0$ is not compatible with a LHV model. Each point's colour represents the estimated amount of noise $p$, from dark blue ($p=1$, {\it i.e.} absence of noise), to light blue ($p=0$, {\it i.e.} maximum noise). Theoretical predictions are shown in the background, the red to yellow colour transition representing the dependence from $p$. Squares (circles) represent those points which violate (don't violate) bilocality inequality \eqref{bilocalineq}.
{\bf b) } Comparison between experimental bilocality violation and maximum CHSH violation in different regimes of noise. Bilocality test is performed with fixed non-optimized measurement settings while CHSH maximum violation is computed applying the Horodecki criterion (Ref. \cite{Horodecki1995}) to a partial quantum state tomography (red points) or a complete quantum state tomography (blue points) of the quantum state shared between A and C after the entanglement swapping protocol (i.e. conditioned on singlet state outcome in station B). The purple point was evaluated directly testing both bilocality and CHSH in a particular regime of low noise. Circles (squares) represent entangled (separable) quantum states, where the degree of entanglement was computed via the partial transpose \cite{Peres1996}. The light blue region is compatible with both models, the light green denotes incompatibility with CHSH, the dark green denotes violation of both bilocality and CHSH inequalities, while the orange region is characterized by only CHSH violation.
} \label{Results2}
\end{figure*}

Another relevant way to visualize the non-bilocal correlations generated in our experiment and its relation to usual local models is displayed in \figref{Results}-b. A bilocal model (defined by \eqref{LHV_bilocal}) must respect the inequality $\sqrt{\vert I \vert}+\sqrt{\vert J \vert}\leq 1$ while a standard LHV model (defined by \eqref{LHV_tripartite}) in turn fulfils $|I|+|J|\leq 1$.  As shown in \figref{Results}-b, the measured values of $I$ and $J$ are clearly incompatible with bilocality (apart from the cases with the highest amount of noise) and behave in good agreement with the theoretical model. Moreover, it clearly shows how optimizing the measurement settings improves the robustness of violation against noise. The data in \figref{Results}-b also shows that the observed values for $I$ and $J$ do not violate the corresponding LHV inequality. However, this only represents a necessary condition. To definitively check whether we are really facing a new type of non-local correlations beyond the standard LHV model \eqref{LHV_tripartite}, we also checked that all Bell inequalities defining our scenario are not violated in the experiment.

In general, given an observed probability distribution, it is a simple linear program to check if it is compatible with LHV model (see e.g. Ref. \cite{Chaves2015b} for further details). Equivalently, noticing that a LHV model defines a polytope of correlations compatible with it \cite{Pitowsky1991}, one can derive all the Bell inequalities constraining that model. As described in the Supplementary Information, we have derived all the Bell inequalities constraining $p(a,b_0,b_1,c \vert x,z)$ compatible with LHV models. Apart from trivial ones, there are $61$ of these inequalities and we have checked for all the collected data with different noise parameter $p$ whether they are violated. The results are shown in \figref{Results2}-a.
It can be seen that all the points (even those that do violate the bilocality inequality \eqref{bilocalineq}, as shown in \figref{Results}) fulfill all LHV constraints, within error bars. It is thus clear that we are facing a new form of non-locality, i.e. non-bilocality.

Finally we addressed the question whether, in an entanglement swapping scenario, bilocality violation could represent a stronger test rather than the usual CHSH violation \cite{Clauser1969}, in order to certify quantum non-local correlations in presence of experimental noise. We therefore performed a tomography of the quantum state shared between Alice and Charlie upon conditioning on Bob's outcome (i.e. entanglement swapped state) followed by an experimental test of bilocality. This allowed us to compare our experimental bilocality violation with the maximum possible CHSH of the swapped state in different regimes of noise \cite{Horodecki1995}. \figref{Results2}-b clearly shows the existence of quantum states which violate bilocality (even without any settings' optimization) but cannot violate the CHSH inequality, thus turning unfeasible any protocol \cite{Ekert1991,Barrett2005,Vazirani2014} based on its violation.\\

Our results provide the first experimental proof-of-principle for network generalizations of Bell's theorem. From a fundamental perspective, recent results \cite{Fritz2012,Oreshkov2012,Spekkens2012,Henson2014,Chaves2015b,Costa2015,Ringbauer2016,Brukner2014,Hoban2015,Ried2015} at the interface between quantum theory and causality have shown that Bell's theorem represents a very particular case of much richer and broader range of phenomena that emerge in complex networks and that hopefully will lead to a deeper understanding of the apparent tension between quantum mechanics and our notions of causal relations. Also, given the close connections between causal inference and machine learning \cite{Spirtes2010}, it is pressing to consider what advantages the recent progresses in quantum machine learning \cite{Wittek2014,Schuld2015} can provide in such a causal context. From a more applied perspective, such generalizations offer an almost unexplored territory. Since network models are more restrictive with respect to classical explanations, they offer a novel route for decreasing the requirements in experimental implementations of nonlocal correlations and thus for their potential applications in the processing of information. For instance, a natural next step is to experimentally realize even larger quantum networks as the one shown in \figref{fig:lhv}-d. For sufficiently long networks, the final quantum state swapped between the end nodes may be separable and thus irrelevant as a quantum resource. Still, the correlations in the entire network might be highly nonlocal \cite{Rosset2016} allowing us to probe a whole new regime in quantum information processing.

\vspace{0.2cm}
\textbf{Acknowledgements}
We thank S. Lloyd for very useful discussions. G.C. thanks Becas Chile and Conicyt for a doctoral fellowship. R.C. acknowledges financial support from the Excellence Initiative of the German Federal and State Governments (Grants ZUK 43 and 81), the FQXi Fund, the US Army Research Office under contracts W911NF-14-1-0098 and W911NF-14-1-0133 (Quantum Characterization, Verification, and Validation), the DFG (GRO 4334 and SPP 1798) and the brazilian ministries MEC and MCTIC. This work was supported by project AQUASIM (Advanced Quantum Simulation and
Metrology).

\section*{Methods}

\small{
{\bf Experimental details}: Photon pairs were generated in two equal parametric down conversion sources, each one composed by a nonlinear crystal (BBO) injected by a pulsed pump field with $ \lambda= 392.5$ nm. The data shown in \figref{Results}, \figref{Results2}-a and the purple point in \figref{Results2}-b  were collected by using 1.5 mm -thick BBO crystals, while for the red and blue points in \figref{Results2}-b, we used 2 mm -thick crystals to increase the generation rate. After spectral filtering and walkoff compensation, photon are sent to the three measurement stations. The observable $A_{0}$, {\it i.e.} $(\sigma_{z}+\sigma_{x})/\sqrt{2}$, corresponds to a HWP rotated by
$\theta^A_0=11.25\si{\degree}$, while $A_{1}$, {\it i.e.} $(\sigma_{z}-\sigma_{x})/\sqrt{2}$, corresponds to $\theta^A_1=78.75\si{\degree}$.  Analogously, $C_{0}$ and $C_{1}$ can be measured at Charlie's station using the same angles $\theta^C_0=\theta^A_0$ and $\theta^C_1=\theta^A_1$.

\vspace{0.4cm}
\textbf{Author contribution}
G.C., L.S., F.A., M.B. and F.S performed the experiment; F.A., G.C., L.S., M.B., R.C. and F.S developed the theoretical tools; all the authors discussed the results and contributed to the writing of the manuscript.

\subparagraph*{Competing financial interests.}
The authors declare no competing financial interest.

\end{document}